\documentstyle{l-aa}
\begin{document}

  \thesaurus{  11          % Galaxies
              (11.01.1;    % Galaxies:abundances
               11.09.4;    % Galaxies: ISM
               11.09.5;    % Galaxies:irregular
               11.17.1) }  % (Galaxies:) quasars: absorption lines
	       
   \title{On the origin of nitrogen in low-metallicity galaxies:}

   \subtitle{Blue compact galaxies versus damped Ly$\alpha$ absorbers}

   \author{ L.S. Pilyugin }

  \offprints{L.S. Pilyugin }

   \institute{   Main Astronomical Observatory
                 of National Academy of Sciences of Ukraine,
                 Goloseevo, 252650 Kiev-22, Ukraine, \\
                 (pilyugin@mao.kiev.ua)}
                 
   \date{Received ; accepted }

\maketitle

\markboth {L.S.Pilyugin: On the nitrogen origin in low-metallicity galaxies}{}

\begin{abstract}

A remarkably small scatter in the N/O ratios for the HII regions in 
low-metallicity blue compact galaxies (BCG) has been found recently. It 
lead to the conclusion that N and O are both produced by massive stars.
Conversely, the N/Si ratios in damped Ly$\alpha$ absorbers (DLA) show 
a large scatter. This result provides support for the time-delay model of 
nitrogen production in intermediate-mass stars, nitrogen production in 
massive stars being not required. The goal of this study is to test whether 
these observational data are compatible with each other and with the existing 
ideas on the chemical evolution of galaxies.

We find that it is possible to reconcile the constancy of N/O ratios in 
low-metallicity BCGs and the scatter of N/Si ratios in DLAs under the 
following three assumptions: {\it 1}) a significant part of nitrogen is 
produced by intermediate-mass stars, {\it 2}) star formation in BCGs and DLAs 
occurs in bursts separated by quiescent periods, {\it 3}) the previous star 
formation events are responsible for measured heavy element abundances of HII 
regions in BCGs.

Since the reality of low N/Si ratios in DLAs is not beyond question, the 
possibility that the nitrogen in low-metallicity BCGs has been  produced by 
massive stars cannot be rejected without additional observations.


\end{abstract}

\section{Introduction}

The theoretical stellar yields of nitrogen are rather uncertain. It has been 
found (Renzini \& Voli 1981; Marigo et al 1996,1998; van den Hoek \& 
Groenewegen 1997) that nitrogen production takes place in the asymptotic 
giant branch phase of the evolution of intermediate-mass stars, but the 
mass range of the nitrogen-producing stars and the predicted 
amount of freshly produced nitrogen depend on poorly known parameters.
The theoretical yields of nitrogen by massive stars are essentially unknown. 
Therefore, an "empirical" approach (analysis of the relation  of N/O 
with O/H) has been widely used to study of the origin of nitrogen
(Edmunds \& Pagel 1978; Lequeux et al 1979; Matteucci \& Tosi 1985; 
Garnett 1990; Pilyugin 1992, 1993; Vila-Costas \& Edmunds 1993; Marconi et al 
1994; among others).

Pagel (1985) called attention to the large scatter in N/O at fixed O/H in  
low-metallicity dwarf galaxies.  Given the time delay between the injection of 
nitrogen by intermediate-mass stars and that of oxygen by shorter lived 
massive stars (the time -- delay hypothesis: Edmunds \& Pagel 1978) and the
hypothesis of self-enrichment of star formation regions (Kunth \& Sargent 1986), 
models for the chemical evolution of dwarf galaxies reproducing the observed 
scatter of N/O have been constructed (Garnett 1990, Pilyugin 1992, Marconi et
al 1994). A significant part (if not all) of nitrogen was assumed in these 
models to be produced by intermediate-mass stars. The HII regions with high 
N/O abundance ratios have been considered to be in early stages of star 
formation (before self-enrichment in oxygen occured), reflecting the chemical 
composition of the whole galaxy. The HII regions in advanced stages of the 
star formation bursts would have small N/O ratios due to self-enrichment in 
oxygen. These HII regions would then reflect the local chemical composition, 
and the scatter in their N/O ratios would be caused by different degrees of 
temporal decrease of N/O ratio within the HII regions. During the interburst 
period the nitrogen is ejected by the intermediate-mass stars, and the matter 
ejected by massive and intermediate-mass stars is supposed to be well mixed 
throughout the whole galaxy.  Some scatter in global N/O ratios among dwarf 
galaxies can be caused by enriched galactic winds (Pilyugin 1993, 1994;
Marconi et al 1994). 

In this framework, the chemical composition of low-metallicity dwarf galaxies 
was expected to be characterised by high N/O ratios with a relatively small 
dispersion, and the large observed dispersion in the N/O ratios is supposed 
to be due to a temporary N/O decrease inside HII regions in which the chemical 
composition is determined spectroscopically.

In contrast with previous measurements, Thuan et al (1995) and Izotov \&
Thuan (1999) have found a remarkably small scatter in the N/O ratios of
the HII regions in low-metallicity (with 12+logO/H $\leq$ 7.6) BCGs. 
Izotov \& Thuan concluded that these galaxies are presently undergoing their 
first burst of star formation, and that nitrogen in these galaxies is produced 
by massive stars only. Since the intermediate and low-mass stars certainly do 
not make an appreciable contribution to oxygen production, the N/O ratio 
corresponding to the ejecta of massive stars and observed in low-metallicity
BCGs is a lower limit for the global N/O ratios. This conclusion is in conflict 
with the fact that the nitrogen to $\alpha$-element abundance ratios measured 
in some DLAs are well below than the typlical value observed in low-metallicity 
BCGs (Lu et al 1998, and references therein). [Since oxygen abundance 
measurements are not available for DLAs,  $[N/S]$ or $[N/Si]$ ratios are
considered instead of $[N/O]$. This is justified by the fact that there is no 
reason to believe that the relative abundances of O, S and Si which are all
produced in Type II supernovae are different from solar in DLAs.] However, as 
suggested by Izotov \& Thuan (1999), the nitrogen to $\alpha$-element abundance 
ratios in DLAs can be significantly underestimated if the absorption lines 
originate in the HII instead of the HI gas. Nevertheless,  the possibility of 
a truly low nitrogen abundances in some DLAs cannot be excluded without 
additional observations, and these DLAs with low nitrogen to $\alpha$-element 
abundance ratios do not confirm the idea that the N/O ratio in low-metallicity 
BCGs is a lower limit for the global N/O ratio in galaxies.

The best way to find the lower limit of N/O ratio and hence the amount of 
nitrogen produced by massive stars would be to determine the N/O ratios in 
galactic halo stars. Unfortunately, at the present state their N/O ratios 
cannot be determined with a precision better than a factor 2 or 3. The only firm 
conclusion is that nitrogen has a strong primary component (Carbon et al 1987).

The goal of this study is to test whether the constancy of the N/O ratios in
low-metallicity BCGs and and the scatter of the N/Si ratios in DLAs are 
compatible with each other and with the existing ideas on the chemical 
evolution of galaxies.

\section{Possible interpretation of nitrogen abundances in BCGs and DLAs}

Here we will demonstrate that the constancy of the N/O ratios in
low-metallicity BCGs and the scatter of the N/Si ratios in DLAs can be 
reconciled under the following three assumptions: {\it 1}) a significant part 
of nitrogen is produced by intermediate-mass stars, {\it 2}) star formation in 
BCGs and DLAs occurs in bursts separated by quiescent periods, {\it 3}) the 
previous star formation events are responsible for heavy element abundances 
observed in the HII regions of the BCGs.

As it was discussed in Introduction, there are no crucial arguments in favor 
of or against assumption {\it 1}. 

Assumption {\it 2} - that star formation in BCGs and DLAs occurs in bursts 
separated by quiescent periods - is commonly accepted in the case of BCGs 
after it was initially suggested by Searle \& Sargent (1972). In order to 
reproduce the observed properties of low-metallicity BCGs, only a few (in some 
cases only one or two) star formation bursts during their life are required 
(Tosi 1994). Papaderos et al (1998) have found that the spectrophotometric 
properties of SBS 0335-052 can be accounted for by a stellar population not 
older than $\sim$ 100 Myr. The possibility of an underlying old (10 Gyr) 
stellar population with mass not exceeding $\sim$ 10 times that of young 
stellar population mass hawever cannot be definitely ruled out on the basis of 
the spectrophotometric properties. In other words, the time interval between 
possible previous and current star formation events can be as large as $\sim$ 
10 Gyr. In the case of DLAs, assumption {\it 2} seems to be also acceptable, 
despite the fact that DLAs do not constitute an homogeneous class of galaxies 
but belong to the wide variety of morphological types of galaxies (Le Brun et 
al 1997). Star formation in a given region in galaxies of any morphological 
type seems to be episodic. The DLA observations sample the general interstellar 
medium at random times along random lines of sight and may or may not see a 
region where a star formation event occured in a recent past.

Assumption {\it 3} - that previous star formation events are responsible for 
the observed heavy element abundances in BCGs - is equivalent to say that the 
element abundances of HII regions in BCGs are not yet polluted by the stars 
of the present star formation event, and that their abundances reflect the 
average N/O in the galaxy, which results from cumulative previous star 
formation. Martin (1996) has found that the current event of star formation in 
the most metal-poor known blue compact galaxy I Zw 18 started 15-27 Myr ago. 
The duration of current star formation burst in another extremely  metal-poor
blue compact galaxy SBS 0335-052 (Papaderos et al 1998) is also in excess 
of the lifetime of the most massive stars. Therefore, a selection effect in 
favor of observations of young HII regions in which the massive stars had 
not yet have time to explode as supernovae cannot be only reason why the HII 
regions are not observed as self-enriched. Massive stars in the current star 
formation burst have often had time to synthesize heavy elements and to eject 
them via stellar winds and supernova explosions into the surrounding 
interstellar gas. Kunth \& Sargent (1986) suggest that the heavy elements 
produced in this way initially mix into H II region only, i.e. the giant H II 
regions are self-enriched. However, it is possible that the nucleosynthetic 
products of massive stars are in high stages of ionization and do not make 
appreciable contribution to the element abundance as derived from optical 
spectra (Kobulnicky \& Skillman 1997, Kobulnicky 1999). Indeed, the oxygen 
abundance in SBS 0335-052 has been measured within the region of 3.6 kpc 
(Izotov et al 1997). There is a supershell of radius $\sim$ 380 pc. There is 
no difference in oxygen abundances inside and outside the supershell as it 
should be expected since $\sim$ 1500 supernovae are required to produce this 
supershell (Izotov et al 1997). Other star-forming galaxies, which are 
chemically homogeneous despite the presence of multiple massive star clusters, 
are reported by Kobulnicky \& Skillman (1998). This can be considered as 
evidence that the nucleosynthetic products of massive stars in giant HII 
regions are hidden from optical spectroscopic searches because they are 
predominantly found in a hot, highly -- ionized superbubble. It should be noted 
however that some fraction of supernova ejecta can mix with dense clouds 
changing their chemical composition. If such cloud survives and produces a 
subgroup of stars shortly, the star formation region will have sub-generations 
of stars with different chemical composition. This seems to be the case in
the Orion star formation region (Cunha \& Lambert 1994, Pilyugin \&
Edmunds 1996).

\begin{figure}[thb]
\vspace{7.0cm}
\includegraphics{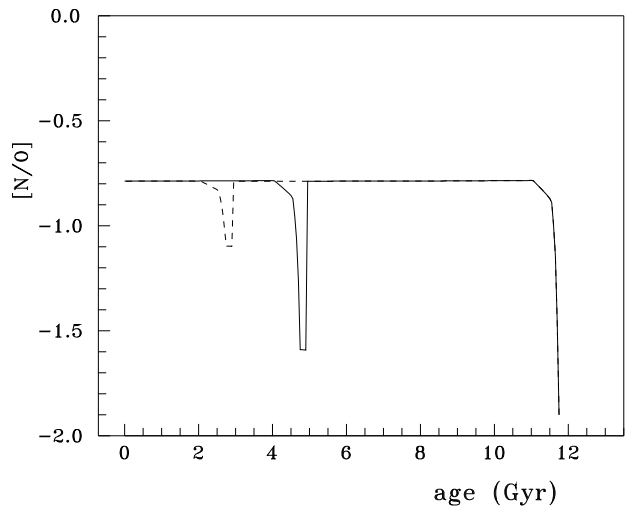}
\caption{\label{figure:8566f1}
The $[N/O]$ as a function of time for two models with different star formation
histories. The dashed line corresponds to the evolution of a system in which 
two star formation bursts of equal intensities occured 12 Gyr and 3 Gyr ago. 
The solid line corresponds to the evolution of a system in which two star 
formation bursts occured 12 Gyr and 5 Gyr ago, the second star formation event 
being 5 times stronger than the first one.}
\end{figure}

With assumptions {\it 1} and {\it 2} the behaviour of the N/O ratio is 
described by models of type suggested by Garnett (1990), Pilyugin (1992), 
Marconi et al (1994). Figgure \ref{figure:8566f1} illustrates the time behaviour 
of N/O in the interstellar medium of a galaxy in the case when all the nitrogen 
is produced by intermediate-mass stars (see also Fig.7 in Pilyugin 1992). Since 
only low-metallicity galaxies are considered here, the adopted total astration 
level in the models is small (less than 0.05), and the nitrogen yield is 
assumed to be independent on metallicity. The nitrogen yields by stars in the 
mass interval 3$\div$4M$_{\odot}$ were taken from Renzini and Voli (1981). 
As can be seen on Fig.\ref{figure:8566f1}, the N/O  ratio increases for about 
1 Gyr after the star formation burst and then remains constant. This is due to 
the assumption that nitrogen is not produced by stars with masses less than 
$\sim$ 3$M_{\odot}$ (Renzini and Voli 1981).  If stars with masses less than 
$\sim$ 3$M_{\odot}$ also make some contribution to the nitrogen production 
(Marigo et al 1996, van den Hoek and Groenewegen 1997), this does not change
appeciably the picture because the amount of nitrogen ejected decreases 
strongly with decreasing stellar mass (Renzini and Voli 1981, Marigo et al 1996, 
van den Hoek and Groenewegen 1997). Therefore, low-metallicity systems with a 
large time interval (more than $\sim$ 1 Gyr) between successive star formation 
events will have close values of N/O ratios  before the current star formation 
event (Fig.\ref{figure:8566f1}).

%============================================================Fig 8566 f1

With this behaviour of the N/O ratio and taking into account assumption {\it 3}, 
the nitrogen abundances in BCGs and DLAs can be interpreted in the following 
way. The low-metallicity BCGs are systems with a small amount of old (with age 
$>$ 1 Gyr) underlying stellar population over which the current star formation 
burst is superposed; only the stars from the previous star formation event(s) 
are responsible for the observed chemical composition in the giant HII regions 
in these galaxies. The DLAs with low nitrogen to $\alpha$-element ratios 
correspond to systems probed less than 1 Gyr after the last local star formation 
event, but after a time sufficient for disappearance of the superbubble and 
mixing of the freshly produced heavy elements in the interstellar medium.
Conversely, the DLAs with nitrogen to $\alpha$-element ratios close to that in 
low-metallicity BCGs correspond to systems in which the time interval after 
last star formation event is sufficiently large for intermediate-mass stars 
to have substantially enhanced the nitrogen to $\alpha$-element abundance ratios.

\section{Discussion and conclusions}

We have shown that the observed nitrogen abundances in both BCGs and DLAs can 
be reproduced if a significant part of nitrogen is produced by intermediate-mass 
stars. If one assumes instead that the nitrogen abundances measured in 
low-metallicity BCGs is produced by massive stars, the observational data for 
BCGs and DLAs are incompatible with each other. Can this be considered as a 
crucial argument in favor of dominant production of nitrogen by 
intermediate-mass stars? Since the low nitrogen to $\alpha$-element abundance 
ratios obtained in some DLAs is not beyond question, the possibility that  
nitrogen in low-metallicity BCGs is produced by massive stars cannot be 
excluded.

The crucial argument in favor of dominant production of nitrogen by 
intermediate-mass stars would be the reliable proof of the existence of 
systems with low N/O ratios. Solid determinations of nitrogen abundances in 
damped Ly$\alpha$ absorbers can clarify this matter. The Zn/S abundance ratios 
in DLAs with low measured nitrogen to $\alpha$-element abundance ratios can also 
tell us something about the reality of low N/O ratios in these objects. The 
measured value of $[Zn/S]$ in DLA associated with QSO 0100-130  is close to 
solar ratio (Prochaska \& Wolfe 1998, Lu et al 1998), indicating that Type I 
supernovae have already contributed to the zink abundance. The measured $[N/S]$ 
in this DLA is close to the $[N/O]$ ratio in low-metallicity BCGs. If values 
of Zn/S in DLAs with low N/O ratios are close to the solar ratio, this will 
be a strong argument in favor of an underestimation of the nitrogen to 
$\alpha$-element abundance ratios. The time scale for nitrogen enrichment is 
shorter than the time scale for iron enrichment, and a system with high Fe/O 
ratio should not have a low N/O ratio. If values of $[Zn/S]$ in DLAs with 
measured low N/O ratios are close to $[Fe/O]$ ratios in the galactic halo stars, 
this can be considered as an argument in favor of genuinely low nitrogen to 
$\alpha$-element abundance ratios in these DLAs. 

Of course, the best way to find the amount of nitrogen produced by  massive
stars would be an undisputable determination of N/O ratios in galactic halo 
stars.

In summary:

If a significant part of nitrogen is produced by intermediate-mass stars, it
is possible to reconcile the observational data for BCGs and DLAs.

If nitrogen is mainly produced by massive stars, the observational data for 
BCGs and DLAs are incompatible with each other. Since the low nitrogen to 
$\alpha$-element abundance ratios obtained in some DLAs are not beyond question,
the possibility that nitrogen measured in low-metallicity BCGs is produced by 
massive stars cannot however be completely excluded.

\begin{acknowledgements}
I would like to thank Drs. N.G.Guseva and Y.I.Izotov for fruitful discussions. 
I thank the referee, Prof. J.Lequeux, for helpful comments and suggestions
which resulted in a better presentation of the work. This work was partly 
supported through INTAS grant 97-0033. This study has been done using the 
NASA's Astrophysical Data Service.
\end{acknowledgements}

\end{document}